\documentclass[reprint,twocolumn]{revtex4}
\usepackage{amsfonts}
\usepackage{amsmath}
\usepackage{amssymb}
\usepackage{charter}
\usepackage{graphicx}
\usepackage{float}

\begin{document}

\title{Simultaneous multiple angular displacement estimation precision
enhanced by the intramode correlation}
\author{Shoukang Chang$^{1}$}
\author{Wei Ye$^{2}$}
\thanks{Corresponding author. 71147@nchu.edu.cn}
\author{Xuan Rao$^{2}$}
\author{Huan Zhang$^{3}$}
\author{Liqing Huang$^{1}$}
\author{Mengmeng Luo$^{4}$}
\author{Yuetao Chen$^{1}$}
\author{Shaoyan Gao$^{1}$}
\thanks{Corresponding author. gaosy@xjtu.edu.cn}
\affiliation{$^{{\small 1}}$\textit{MOE Key Laboratory for Nonequilibrium Synthesis and
Modulation of Condensed Matter, Shaanxi Province Key Laboratory of Quantum
Information and Quantum Optoelectronic Devices, School of Physics, Xi'an
Jiaotong University, 710049, People's Republic of China}\\
$^{{\small 2}}$\textit{School of Information Engineering, Nanchang Hangkong
University, Nanchang 330063, China}\\
$^{{\small 3}}$\textit{School of Physics, Sun Yat-sen University, Guangzhou
510275, China}\\
$^{4}$\textit{Department of Physics, Xi'an Jiaotong University City College,
Xi'an 710018, China}}

\begin{abstract}
The angular displacement estimation is one of significant branches of
quantum parameter estimation. However, most of the studies have focused on
the single-angular displacement estimation, while the multiple angular
displacement estimation in ideal and noisy scenarios is still elusive. In
this paper, we investigate the simultaneous multiple angular displacement
estimation based on an orbital angular momentum (OAM), together with
inputting $(d+1)$-mode NOON-like states as the probe state. By revealing the
role of the intramode correlation of the probe state, this allows us to give
a reasonable explanation for the corresponding quantum Cram\'{e}r-Rao bound
(QCRB) behaviors with and without photon losses. Our analyses suggest that
the QCRB for the multiple angular displacement estimation is always
positively related to the intramode correlation, especially for the
multimode entangled squeezed vacuum state showing the best performance
compared to another probe state. More importantly,\ strengthening the
robustness of multiple angular-displacement estimation systems can be
achieved by increasing the OAM quantum number.

\textbf{PACS: }03.67.-a, 05.30.-d, 42.50,Dv, 03.65.Wj
\end{abstract}

\maketitle

\section{Introduction}

Quantum parameter estimation provides a feasible way to more accurately
estimate\ physical quantities that can not be measured directly than its
classical counterpart \cite{1,2,3,4}. As a specific example, in
phase-estimated systems, the usage of quantum resources, involving
nonclassical and entanglement states, can make the phase sensitivity beat
the so-called shot-noise limit, even closing to the renowned Heisenberg
limit \cite{5,6,7}. In general, the precision limit of quantum parameter
estimation can be visually quantified by means of the quantum Cram\'{e}r-Rao
bound (QCRB), which is not only inversely proportional to the quantum Fisher
information (QFI) \cite{2,8}, but also has been extensively studied and used
especially in quantum single-(or multi-) phase estimation.

Originally, a conventional model to study the quantum parameter estimation
is the phase estimation problem \cite{9}. In particular, taking advantage of
optical interferometers, such as a Mach--Zehnder interferometer \cite%
{10,11,12} and an SU(1,1) interferometer \cite{13,14,15,16}, early
investigations of phase estimation pay attention to the single-phase
estimation since it can be easily realized both theoretically and
experimentally \cite{5,11,15}. More strikingly, the single-phase estimation
with the QCRB in the presence of noisy environments, e.g., photon loss \cite%
{17,18,19}, phase diffusion \cite{20,21}, and thermal noise \cite{22,23},
can be tackled using the variational method \cite{17,20} proposed by Escher,
greatly promoting the practical applications of quantum metrology \cite%
{24,26,27}. On the other hand, extending toward the multiple phase
estimation\ with the QCRB has attracted considerable interest more recently,
thereby resulting in the potential applications \cite%
{25,28,29,30,31,32,33,34}, such as quantum-enhanced sensor network \cite%
{29,30,31,32} and optical imaging \cite{33,34}. Moreover, in order to
improve the precision of multiple-phase estimation, multimode NOON (or
NOON-like) states \cite{35,36,37,38,39}, generalized entangled coherent
states \cite{40} and multimode Gaussian states \cite{41} have been
considered, even in the presence of noisy environment \cite{42,43,44,45}.
More interestingly, by using correlated quantum states, the simultaneous
estimation performance of multiple phases can show a significant advantage
scaling as $O(d)$ with the number of phase shifts $d$ over the optimal
individual case \cite{35}, but the $O(d)$ advantage would fade away in
photon-loss scenarios \cite{45}. Further, in order to find saturable QCRB in
multiparameter estimation, the necessary and sufficient conditions for
projective measurements to saturate the QFI for multiple phase estimation
with pure probe states can be achieved \cite{46}.

In addition to the phase-estimated systems, the angular displacement
estimation based on an orbital angular momentum (OAM) has been one of
important branches of parameter-estimated systems, particularly when the OAM
quantum number $l$ that is theoretically unbounded can give rise to the
unbounded increase in the estimation precision \cite{47,48,49}. Although the
OAM values as high as 10010 quanta have been proven experimentally \cite{50}%
, this value is not indeed unbounded via the limited aperture of optical
systems \cite{47,50,51}. As a result, other methods have to be found to
improve the angular displacement estimation. For instance, to show the
increased performance of angular displacement estimation, the usages of
entangled photon states \cite{49} and twisted N00N states \cite{47} were
taken into account. Apart from the aforementioned methods of generating the
probe states, Maga\~{n}-Loaiza \emph{et al }presented the quantum-improved
sensitive estimation of angular rotations based on a sort of weak-value
amplification \cite{52}. More dramatically, in ideal and realistic
scenarios, Zhang \emph{et al }suggested a super-resolved angular
displacement estimation protocol using a Sagnac interferometer together with
parity measurement \cite{53}. Even so, it should be noticed that these
studies mentioned above pay attention to the single-angular displacement
estimation systems, whereas the multiple angular displacement estimation
problem in the ideal and noisy environments has not been studied before.
Therefore, in this paper, we shall present the derivation of the QCRB for
the multiple angular displacement estimation with and without the photon
losses when using the $(d+1)$-mode NOON-like states [including the multimode
NOON state (MNOONS), the multimode entangled coherent state (MECS), the
multimode entangled squeezed vacuum state (MESVS) and the multimode
entangled squeezed coherent state (MESCS)] as the probe states. Our results
find that, the QCRB for the multiple angular displacement estimation in both
ideal and photon-loss cases is positively associated with the intramode
correlation, especially for the MESVS exhibiting the best performance when
comparing to other probe states. More interestingly,\ the OAM quantum number
$l$ can be profitably used for strengthening the robustness of multiple
angular displacement estimation systems.

The rest of this paper is arranged as follows. In Sec. 2, we first describe
the QCRB for the multiple angular displacement estimation with $d$
independent angular displacements in the ideal scenario, and then focus on
the behaviors of the QCRB when given the four specific probe states. In Sec.
3, we consider the effects of photon losses on the multiple angular
displacement estimation precision, and also analyze its QCRB with the four
probe states under the photon losses. Finally, conclusions are presented in
the last section.

\section{The QCRB for the multiple angular displacement estimation in the
ideal scenario}

In an ideal case, let us beginning with the description of the QCRB for the
simultaneous estimation with $d$ independent angular displacements, whose
schematic diagram is shown in Fig. 1. To be more specific, here we first
take a balanced $(d+1)$-mode entangled pure as the probe state, which can be
defined as \cite{36}
\begin{equation}
\left \vert \Psi \right \rangle =\breve{N}\underset{m=0}{\overset{d}{\sum }}%
\left \vert 0\right \rangle _{0}\left \vert 0\right \rangle _{1}\left \vert
0\right \rangle _{2}...\left \vert \psi \right \rangle _{m}...\left \vert
0\right \rangle _{d},  \label{1}
\end{equation}%
where $\breve{N}=[(1+d)(1+d\left \vert \left \langle \psi |0\right \rangle
\right \vert ^{2})]^{-1/2}$ is the normalization factor. According to Eq. (%
\ref{1}), it is obvious that this probe state is a superposition of $d+1$
multimode quantum states with both an arbitrary single-mode quantum state $%
\left \vert \psi \right \rangle _{m}$ on the $m$th mode and a zero photon
state on the other modes. It should be mentioned that, when $\left \vert
\psi \right \rangle _{m}$ is respectively the Fock state $\left \vert
N\right \rangle _{m}$, the coherent state $\left \vert \alpha \right \rangle
_{m}$, the squeezed vacuum state $\left \vert r_{1}\right \rangle _{m}$ and
the squeezed coherent state $\left \vert \beta ,r_{2}\right \rangle _{m}$,
one can obtain the MNOONS $\left \vert \Psi _{N}\right \rangle $, the MECS $%
\left \vert \Psi _{\alpha }\right \rangle $, the MESVS $\left \vert \Psi
_{r_{1}}\right \rangle $ and the MESCS $\left \vert \Psi _{\beta
,r_{2}}\right \rangle $, which will be seen as the probe state to analyze
the behaviors of the QCRB in the following sections.
\begin{figure}[tbp]
\label{Fig1} \centering \includegraphics[width=0.9\columnwidth]{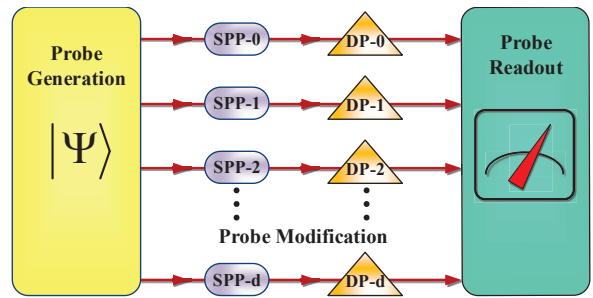}%
\newline
\caption{{} Schematic diagram of multiple angular displacement estimation
with $d$ angular displacements, where a given probe state $\left \vert \Psi
\right \rangle $ after passing through spiral phase plate (SPP) and Dove
prisms (DP) with the same number $d+1$ can be in readout. }
\end{figure}

Subsequently, the generated probe state $\left \vert \Psi \right \rangle $
is sent to $d+1$ spiral phase plates (SPPs) which introduce the OAM degree
of freedom, and after undergoing the $d+1$ Dove prisms (DPs) to generate $d$
independent angular displacements $\theta _{m}$ to be estimated (here $%
\theta _{0}=0$ is viewed as the reference beam), the corresponding evolution
operator can be expressed as%
\begin{equation}
\hat{U}_{\theta }=\exp \left( i\sum_{m=1}^{d}2l\hat{n}_{m}\theta _{m}\right)
,  \label{2}
\end{equation}%
where $l$ is the quanta number of the OAM, $\hat{n}_{m}=\hat{a}_{m}^{\dagger
}\hat{a}_{m}$ and $\theta _{m}$ denote the photon number operator and the
angular displacement on mode $m,$ respectively. After the interaction
between the probe state and the evolution operator $\hat{U}_{\theta },$ the
resulting state becomes $\left \vert \Psi _{\theta }\right \rangle =\hat{U}%
_{\theta }\left \vert \Psi \right \rangle ,$ so that the QCRB for the
multiple angular displacement estimation in an ideal scenario is given by
\cite{35,36,37,38}

\begin{equation}
\left \vert \delta \theta \right \vert ^{2}\geq \left \vert \delta \theta
\right \vert _{QCRB}^{2}=\text{Tr}\left( F^{-1}\right) ,  \label{3}
\end{equation}%
where $F^{-1}$ represents the inverse matrix of the $d\times d$ quantum
Fisher information matrix (QFIM). Generally speaking, the QCRB for the
multiple angular displacement estimation is not achievable. Nevertheless,
for the unitary angular displacement process, i.e., $\left \vert \Psi
_{\theta }\right \rangle =\hat{U}_{\theta }\left \vert \Psi \right \rangle ,$
the QCRB of pure quantum states can be saturated if the probe state $%
\left
\vert \Psi \right \rangle $ satisfies \cite{40,54}
\begin{eqnarray}
&&\left \langle \Psi \right \vert \left[ i(\partial \hat{U}_{\theta
}^{\dagger }/\partial \theta _{j})\hat{U}_{\theta },i(\partial \hat{U}%
_{\theta }^{\dagger }/\partial \theta _{m})\hat{U}_{\theta }\right] \left
\vert \Psi \right \rangle  \notag \\
&=&\left \langle \Psi \right \vert \left[ 2l\hat{n}_{j},2l\hat{n}_{m}\right]
\left \vert \Psi \right \rangle  \notag \\
&=&0,\text{ }\forall j,m,  \label{4}
\end{eqnarray}%
where $\hat{n}_{j,m}$ are the photon number operators on modes $j$ and $m$.
Since both $\hat{n}_{j}$ and $\hat{n}_{m}$ are the Hermitian and mutually
commuting operators, i.e., $\left[ \hat{n}_{j},\hat{n}_{m}\right] =0,$ $%
\forall j,m,$ it is easy for the probe state $\left \vert \Psi
\right
\rangle $ to find that the saturation condition is always true.
Thus, its elements of the QFIM can be given by

\begin{equation}
F_{jm}=16l^{2}\text{Cov}(\hat{n}_{j},\hat{n}_{m}),  \label{5}
\end{equation}%
where Cov$(\hat{n}_{j},\hat{n}_{m})=\left \langle \hat{n}_{j}\hat{n}%
_{m}\right \rangle -\left \langle \hat{n}_{j}\right \rangle \left \langle
\hat{n}_{m}\right \rangle $ is the covariance between the photon number
operators $\hat{n}_{j}$ and $\hat{n}_{m},$ and the average $\left \langle
\cdot \right \rangle $ is taken with respect to the probe state $\left \vert
\Psi \right \rangle $. Combining Eqs. (\ref{1}) and (\ref{5}), as a result,
the QFIM can be calculated as%
\begin{equation}
F=16l^{2}\left[ \left \langle \hat{n}_{m}^{2}\right \rangle I-\left \langle
\hat{n}_{m}\right \rangle ^{2}\tilde{I}\right] ,  \label{6}
\end{equation}%
where $I$ is the $d\times d$ identity matrix and $\tilde{I}$ represents the
matrix with the elements $\tilde{I}_{jm}=1,$ for all $j$ and $m$. Upon
substituting Eqs. (\ref{6}) into (\ref{3}), the analytical expression of the
QCRB for the multiple angular displacement estimation with the probe state $%
\left \vert \Psi \right \rangle $ shown in Eq. (\ref{1}) can be finally
derived by
\begin{eqnarray}
&&\left \vert \delta \theta \right \vert _{QCRB}^{2}  \notag \\
&=&\frac{d}{16l^{2}(\bar{n}_{m}^{2}g_{m}^{\left( 2\right) }+\bar{n}_{m})}%
\left( 1+\frac{1}{g_{m}^{\left( 2\right) }+\bar{n}_{m}^{-1}-d}\right) ,
\label{7}
\end{eqnarray}%
where $\bar{n}_{m}=\left \langle \hat{n}_{m}\right \rangle $ denotes the
average photon number of the probe state $\left \vert \Psi \right \rangle $
on mode $m$, and $g_{m}^{\left( 2\right) }=$ $\left \langle \hat{a}%
_{m}^{\dagger }\hat{a}_{m}^{\dagger }\hat{a}_{m}\hat{a}_{m}\right \rangle /%
\bar{n}_{m}^{2}$ is the second-order coherence function, represented as an
intramode correlation \cite{55}. Generally speaking, the smaller the value
of the QCRB, the more precise the parameter estimation. According to Eq. (%
\ref{7}), notably, the QCRB is positively correlated with the intramode
correlation $g_{m}^{\left( 2\right) }$. That is to say, the intramode
correlation contributes to the enhancement of multiple angular displacement
estimation precision.
\begin{figure}[tbp]
\label{Fig2} \centering \includegraphics[width=0.72\columnwidth]{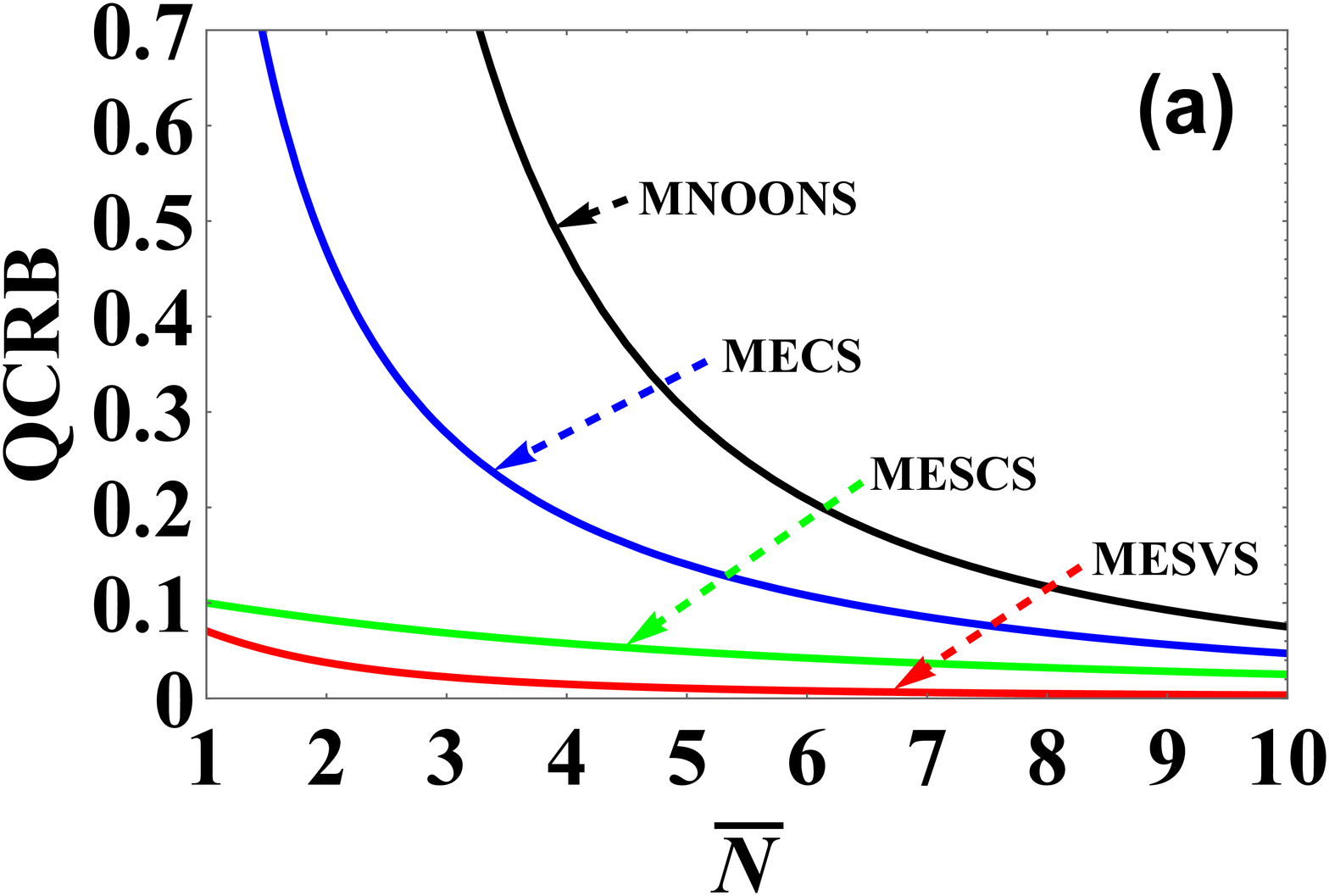}%
\newline
\includegraphics[width=0.72\columnwidth]{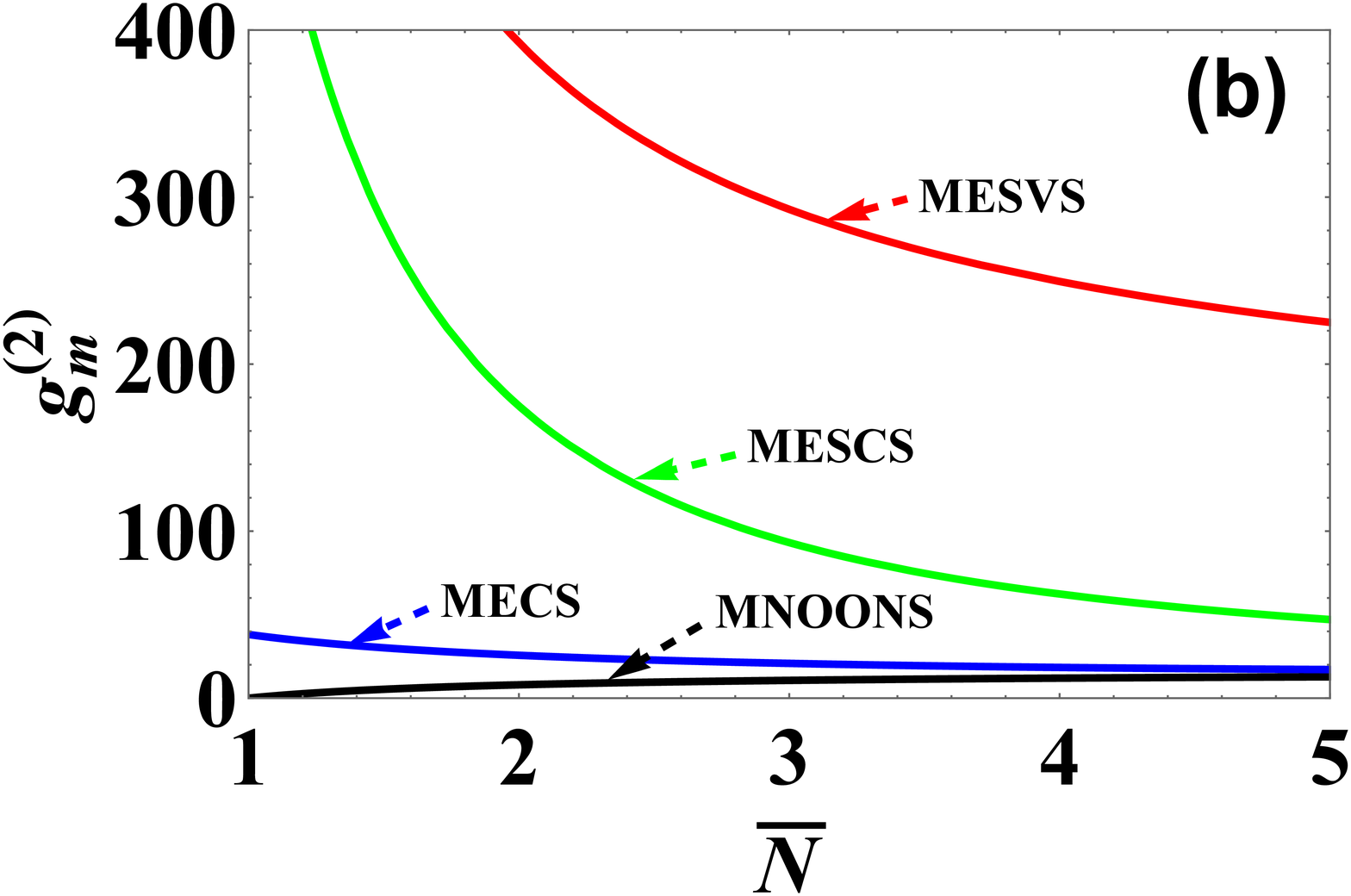} \newline
\caption{{}(Color online) Both (a) the QCRB and (b) the second-order
coherence function $g_{m}^{\left( 2\right) }$ for the multiple angular
displacement estimation as a function of the total mean photon number $\bar{N%
}$ with several different probe states, i.e., the MNOONS (black line), the
MECS (blue line), the MESVS (red line), and the MESCS (green line), at fixed
parameters of $l=2$ and $d=15$.}
\end{figure}

\begin{figure}[tbp]
\label{Fig3} \centering \includegraphics[width=0.72\columnwidth]{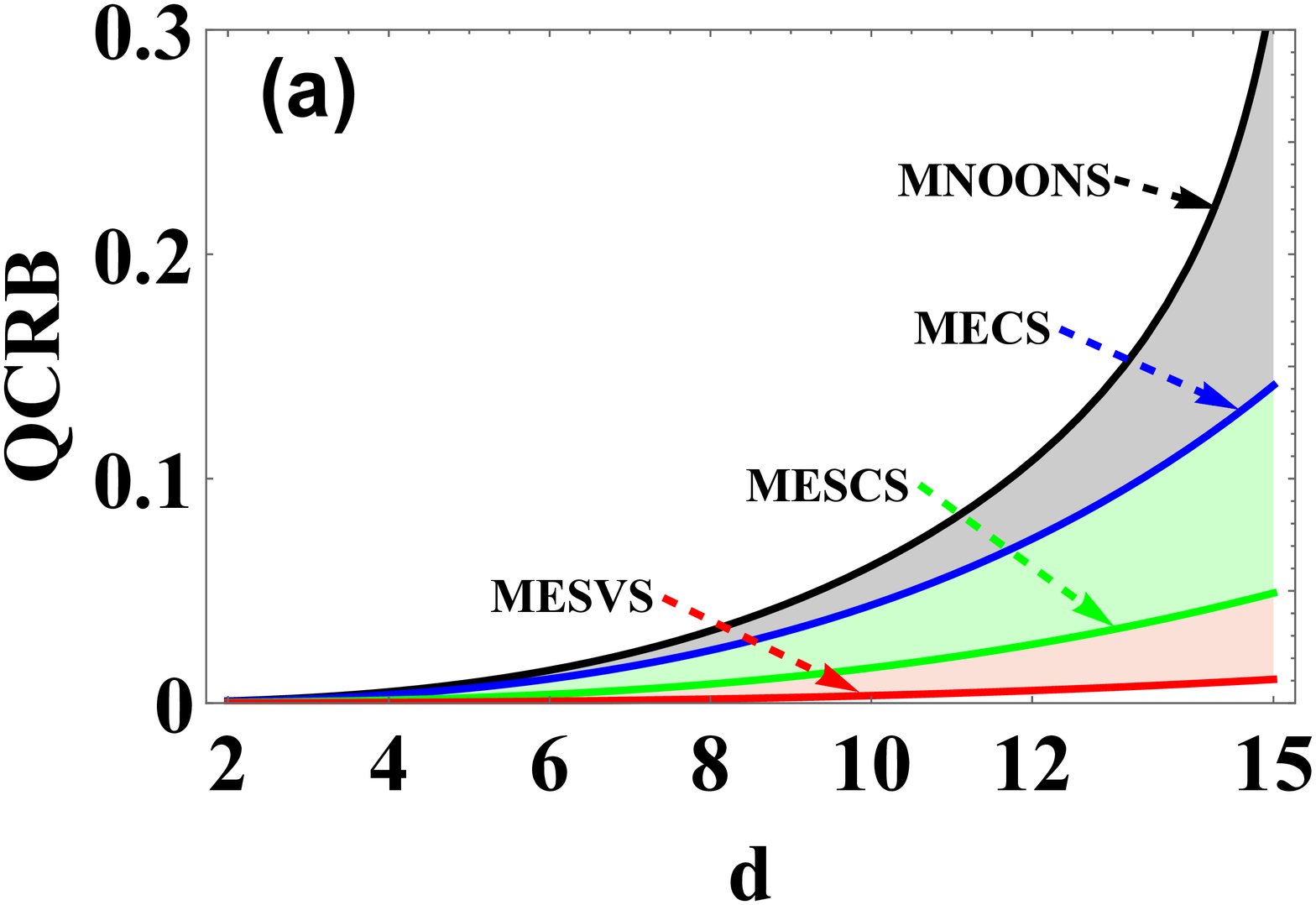}%
\newline
\includegraphics[width=0.72\columnwidth]{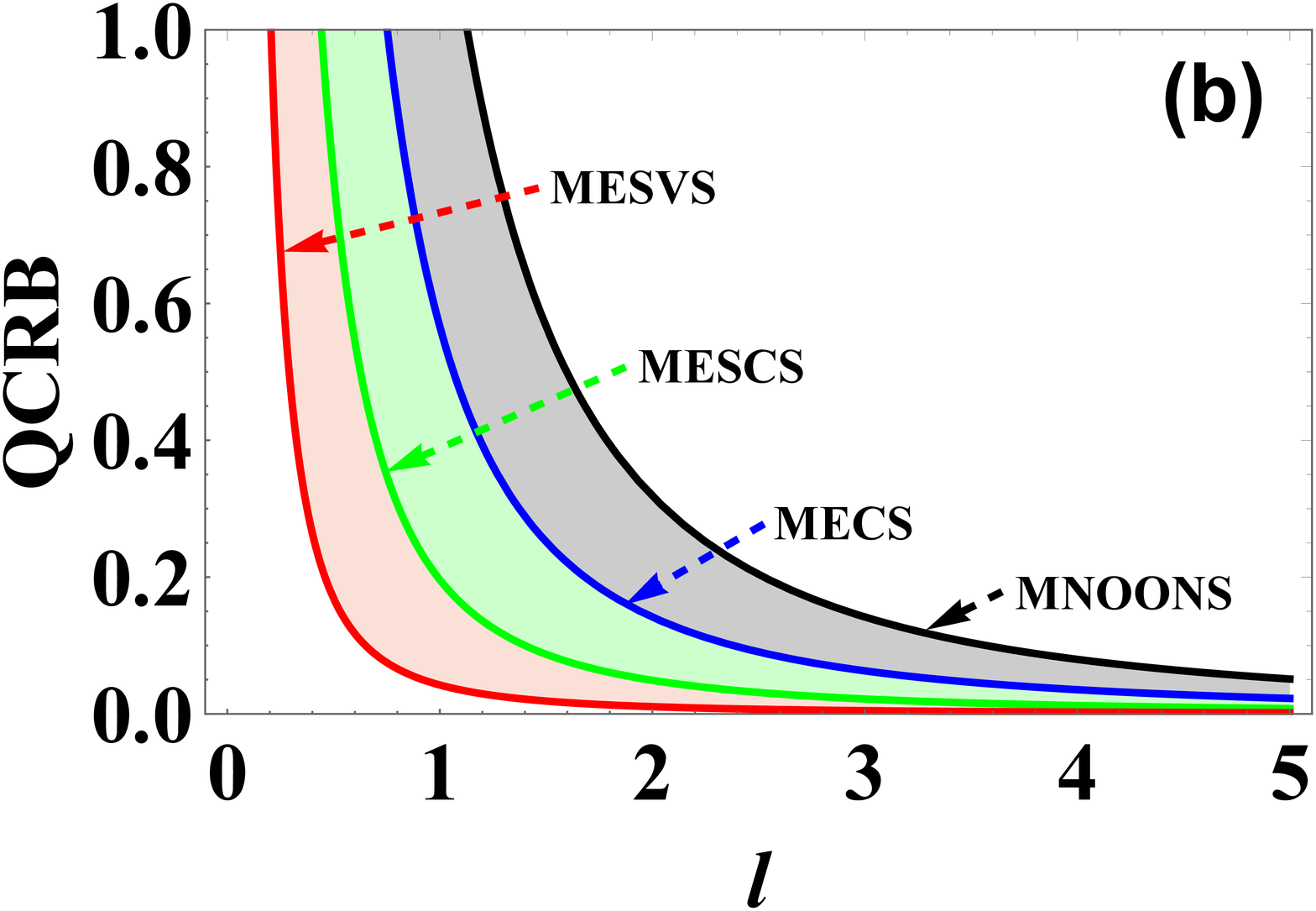} \newline
\caption{{}(Color online) The QCRB for the multiple angular displacement
estimation as a function of (a) the independent angular-displacement number $%
d$ with $l=2$ and $\bar{N}=5,$ and of (b) the quanta number of the OAM $l$
with $d=15$ and $\bar{N}=5,$ when given several different probe states,
i.e., the MNOONS (black line), the MECS (blue line), the MESVS (red line),
and the MESCS (green line). }
\end{figure}
To clearly see the behaviors of the QCRB for the multiple angular
displacement estimation, here we take four specific probe states into
account, including the MNOONS $\left \vert \Psi _{N}\right \rangle $, the
MECS $\left \vert \Psi _{\alpha }\right \rangle $, the MESVS $\left \vert
\Psi _{r_{1}}\right \rangle $, and the MESCS $\left \vert \Psi _{\beta
,r_{2}}\right \rangle $ as the probe states [see Appendix A for more
details]. Without loss of generality, we also assume that both the amplitude
$\alpha $ ($\beta $) of coherent states and the squeezing parameter $r_{1}$ (%
$r_{2}$) are real numbers, so as to achieve the total mean photon numbers $%
\bar{N}$ for the above four multimode entangled states [see Eq. (10) in Ref.
\cite{36}]. In this case, Fig. 2(a) shows the QCRB for the four multimode
entangled states changing with the total mean photon number $\bar{N}$, when
fixed values of $l=2$ and $d=15$. It is shown that the value of the QCRB for
the given multimode entangled states rapidly decreases with the increase of $%
\bar{N}$. Moreover, at the same total mean photon number $\bar{N},$ the
MESVS (red line) shows the lowest QCRB value, followed by the MESCS (green
line), the MECS (blue line) and the MNOONS (black line), which means that
the usage of the MESVS as the probe state can achieve the highest estimation
precision. The reason for this phenomenon is that the intramode correlation
of the MESVS is the strongest in comparison to another multimode probe
state, as shown in Fig. 2(b). In this sense, it is also demonstrated that
the intramode correlation is conducive to effectively improve the multiple
angular displacement estimation precision.

On the other hand, we also consider the effects of both the number of
independent angular displacements $d$ and the quanta number of the OAM $l$
on the QCRB, as depicted in Fig. 3. It is clearly seen from Fig. 3(a) that\
when fixed parameters of $l=2$ and $\bar{N}=5$, the QCRB for the four probe
states increases with the increase of $d$, meaning that as the number of
independent angular displacements $d$ increases, the multiple angular
displacement estimation precision becomes worse. This phenomenon results
from that the QCRB is passively correlated with the number of independent
angular displacements $d$, as given in Eq. (\ref{7}). Even so, as we can see
in Fig. 3(b), at fixed parameters of $d=15$ and $\bar{N}=5$, when increasing
the quanta number of the OAM $l$, the QCRB for the four probe states tends
to be smaller and smaller. This reflects, to some extent, that increasing $l$
can effectively improve the multiple angular displacement estimation
precision. More importantly, it is seen from Fig. 3 that compared to other
probe states, the MESVS still maintains the highest estimation precision.

\begin{figure}[tbp]
\label{Fig4} \centering \includegraphics[width=0.9\columnwidth]{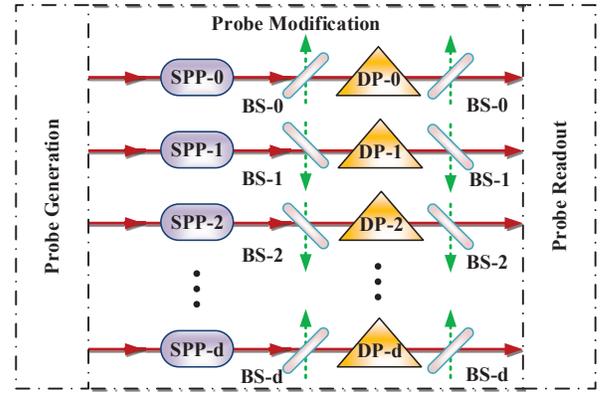}
\caption{{}(Color online) Schematic diagram of multiple angular displacement
estimation with $d$ angular displacements under the photon losses occurring
at both ends of $d+1$ DPs. Here we use the fictitious beam splitter (BS)
with a transmissivity $\protect \eta _{m}$ to simulate a photon-loss process }
\end{figure}

\section{\protect \bigskip The QCRB for the multiple angular displacement
estimation with photon losses}

In the real-life scenarios, the inevitable interaction between the probe
state system $S$ and its surrounding environment $E$ is always existed,
greatly making the parameter-estimated performance worse. In general, there
are various interactions, such as photon loss, phase diffusion, and thermal
noise. For the sake of simplicity, here we\ only pay attention to how the
photon losses affect the multiple angular displacement estimation precision.
In addition, it should be noted that the probe state interacts with the $d+1$
DPs to generate $d$ independent angular displacements $\theta _{m}$ in the
photon-loss environment, which would no longer be an unitary evolution. This
also leads to that, for the multiple angular displacement estimation with
photon losses, the methods used to derive the QCRB given in Eq. (\ref{7})
can not be directly employed. Fortunately, with the assistance of an
variational method, Yue \emph{et al}\textit{. }derived the general form of
the QCRB of multiphase estimation systems in the photon-loss case \cite{45}.
By extending that work \cite{45}, in this section, we shall utilize the
variational method to study the effects of photon losses on the multiple
angular displacement estimation precision (see Fig.4), such that a brief
review of this variational approach is necessary in the following.

When given an initial $(d+1)$-mode probe state $\left \vert \Psi
\right \rangle _{S}$ in the probe system $S$ and an initial state $\left \vert
\vec{0}\right \rangle _{E}$ in the photon-loss environment, it is essential
to expand the sizes of both the probe system space $S$ and the photon-loss
environment space $E,$ thereby resulting in that the probe state $\left \vert
\Psi \right \rangle _{S}$ in the enlarged system-environment space $S+E$
undergoes the unitary evolution $\hat{U}_{S+E}(\theta ),$ which can be
expressed as \cite{45}
\begin{eqnarray}
&&\left \vert \Psi (\theta )\right \rangle _{S+E}  \notag \\
&=&\hat{U}_{S+E}(\theta )\left \vert \Psi \right \rangle _{S}\left \vert \vec{0}%
\right \rangle _{E}  \notag \\
&=&\sum_{k}\hat{\Pi}_{k}(\theta )\left \vert \Psi \right \rangle
_{S}\left \vert \vec{k}\right \rangle _{E},  \label{8}
\end{eqnarray}%
where $\hat{U}_{S+E}(\theta )=\otimes _{m=0}^{d}\hat{U}_{S+E}^{m}(\theta
_{m})$ is the unitary evolution operator, $\left \vert \vec{0}\right \rangle
_{E}=$ $\otimes _{m=0}^{d}\left \vert 0\right \rangle _{E_{m}}$ is the initial
state of environment, $\left \vert \vec{k}\right \rangle _{E}$ $=\otimes
_{m=0}^{d}\left \vert k_{m}\right \rangle _{E_{m}}$is the orthogonal basis of
the environment, and $\hat{\Pi}_{k}(\theta )=\otimes _{m=0}^{d}\hat{\Pi}%
_{k_{m}}(\theta _{m})$ is the direct product of all kraus operator, defined
as%
\begin{eqnarray}
&&\hat{\Pi}_{k_{m}}(\theta _{m})  \notag \\
&=&\sqrt{\frac{(1-\eta _{m})^{k_{m}}}{k_{m}!}}e^{i2l\theta _{m}(\hat{n}%
_{m}-\delta _{m}k_{m})}\eta _{m}^{\frac{\hat{n}_{m}}{2}}\hat{a}_{m}^{k_{m}},
\label{9}
\end{eqnarray}%
with the variational parameters $\delta _{m}$ ($\delta _{m}=0$ and $-1$ are
respectively the photon losses occurring before and after the $d+1$ DPs),
and $\eta _{m}$ quantifying the strength of the photon losses. In practice,
such a photon-loss strength can be often regarded as the transmissivity of
fictitious beam splitters, as seen in Fig. 4. Among them, $\eta _{m}=0$ and $%
1$ respectively indicate the complete-absorption and lossless cases. In this
situation, the QCRB for the multiple angular displacement estimation under
the photon losses turns out to be \cite{45}
\begin{equation}
\left \vert \delta \theta \right \vert _{QCRB_{L}}^{2}=\max_{\hat{\Pi}%
_{k}(\theta )}\text{Tr}[C_{Q}^{-1}(\theta ,\hat{\Pi}_{k}(\theta ))],
\label{10}
\end{equation}%
where $C_{Q}(\theta ,\hat{\Pi}_{k}(\theta ))$ is the QFIM for the enlarged
system-environment space $S+E$, and the matrix elements of $C_{Q}(\theta ,%
\hat{\Pi}_{k}(\theta ))$ are expressed as%
\begin{equation}
C_{Q_{jm}}(\theta ,\hat{\Pi}_{k}(\theta ))=4\left[ \left \langle \hat{\Lambda}%
_{jm}\right \rangle -\left \langle \hat{\Gamma}_{j}\right \rangle \left \langle
\hat{\Gamma}_{m}\right \rangle \right] ,  \label{11}
\end{equation}%
with%
\begin{align}
\hat{\Gamma}_{m}& =i\sum \limits_{k_{m}}\frac{d\hat{\Pi}_{k_{m}}^{\dagger
}(\theta _{m})}{d\theta _{m}}\hat{\Pi}_{k_{m}}(\theta _{m}),  \notag \\
\hat{\Lambda}_{jm}& =\left \{
\begin{array}{c}
\sum \limits_{k_{m}}\frac{d\hat{\Pi}_{k_{m}}^{\dagger }(\theta _{m})}{d\theta
_{m}}\frac{d\hat{\Pi}_{k_{m}}(\theta _{m})}{d\theta _{m}},\text{ }j=m \\
\hat{\Gamma}_{j}\hat{\Gamma}_{m},\text{ }j\neq m%
\end{array}%
\right. .  \label{12}
\end{align}%
Upon substituting Eqs. (\ref{9}) into (\ref{12}), one can further obtain%
\begin{align}
\hat{\Gamma}_{m}& =2l\chi _{m}\hat{n}_{m},  \notag \\
\hat{\Lambda}_{jm}& =\left \{
\begin{array}{c}
4l^{2}(\chi _{m}^{2}\hat{n}_{m}^{2}+\gamma _{m}\hat{n}_{m}),\text{ }j=m \\
\hat{\Gamma}_{j}\hat{\Gamma}_{m},\text{ }j\neq m%
\end{array}%
\right. ,  \label{13}
\end{align}%
with $\chi _{m}=1-\left( 1+\delta _{m}\right) \left( 1-\eta _{m}\right) $
and $\gamma _{m}=\eta _{m}(1-\eta _{m})\left( 1+\delta _{m}\right) ^{2}$.
For the sake of calculation, here we only consider the specific cases of $%
\eta _{m}=\eta $ and $\delta _{m}=\delta $ for any $m$. Thus, based on Eqs. (%
\ref{11}) and (\ref{13}), one can derive the lower bound of the QCRB for the
multiple angular displacement estimation, i.e.,%
\begin{eqnarray}
&&\text{Tr}[C_{Q}^{-1}]  \notag \\
&=&\frac{(d-1)\breve{N}^{-2}}{16l^{2}\sigma }+\frac{\breve{N}^{-2}}{16l^{2}%
\left[ \sigma -d\breve{N}^{2}\chi ^{2}\left \langle \psi \right \vert \hat{n}%
\left \vert \psi \right \rangle ^{2}\right] },  \label{14}
\end{eqnarray}%
where $\sigma =\chi ^{2}\left \langle \psi \right \vert \hat{n}^{2}\left \vert
\psi \right \rangle +\gamma \left \langle \psi \right \vert \hat{n}\left \vert
\psi \right \rangle .$ To further simplify the calculation, we also assume
that $d\gg 1,$ leading to that the second term is infinitesimal compared
with the first term given in Eq. (\ref{14}), such that%
\begin{equation}
\text{Tr}[C_{Q}^{-1}]\approx \frac{(d-1)\breve{N}^{-2}}{16l^{2}\sigma }.
\label{15}
\end{equation}%
In order to maximize Tr$[C_{Q}^{-1}]$, the optimal value of $\delta $ can be
easily calculated as%
\begin{eqnarray}
&&\delta _{opt}  \notag \\
&=&\frac{\left \langle \psi \right \vert \hat{n}^{2}\left \vert \psi
\right \rangle }{\left( 1-\eta \right) \left \langle \psi \right \vert \hat{n}%
^{2}\left \vert \psi \right \rangle +\eta \left \langle \psi \right \vert \hat{n}%
\left \vert \psi \right \rangle }-1.  \label{16}
\end{eqnarray}%
Therefore, substituting Eqs. (\ref{16}) into (\ref{15}), one can obtain the
explicit expression of the QCRB for the multiple angular displacement
estimation in the presence of photon losses, i.e.,%
\begin{eqnarray}
&&\left \vert \delta \theta \right \vert _{QCRB_{L}}^{2}  \notag \\
&=&\frac{d-1}{16l^{2}\bar{n}_{m}}\left( \frac{1-\eta }{\eta }+\frac{1}{1+%
\bar{n}_{m}g_{m}^{\left( 2\right) }}\right) .  \label{17}
\end{eqnarray}%
From Eq. (\ref{17}), it is clear that the QCRB is also positively correlated
with the intramode correlation $g_{m}^{\left( 2\right) }$ even in the
presence of photon losses.
\begin{figure}[tbp]
\label{Fig5} \centering \includegraphics[width=0.72\columnwidth]{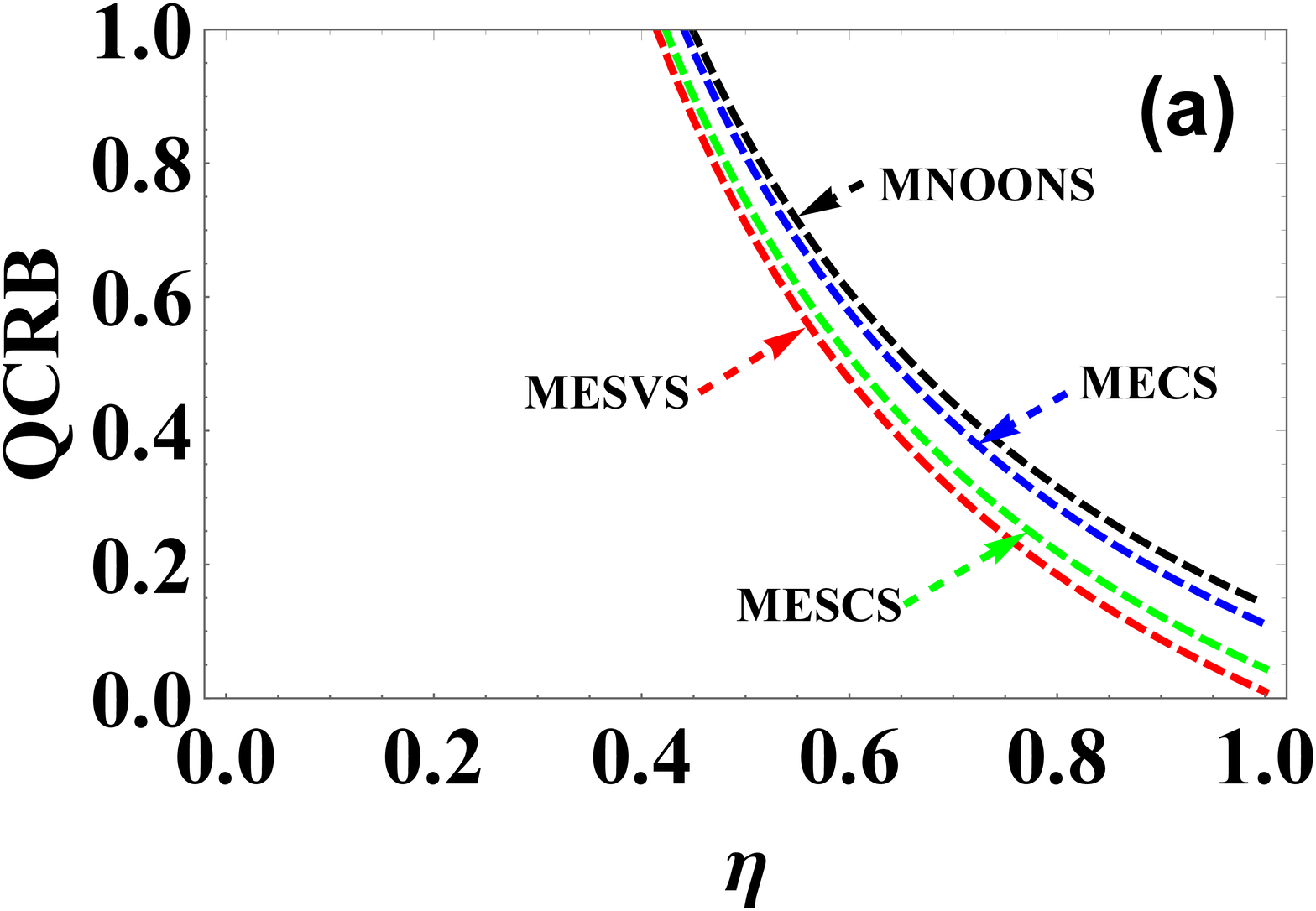}%
\newline
\includegraphics[width=0.72\columnwidth]{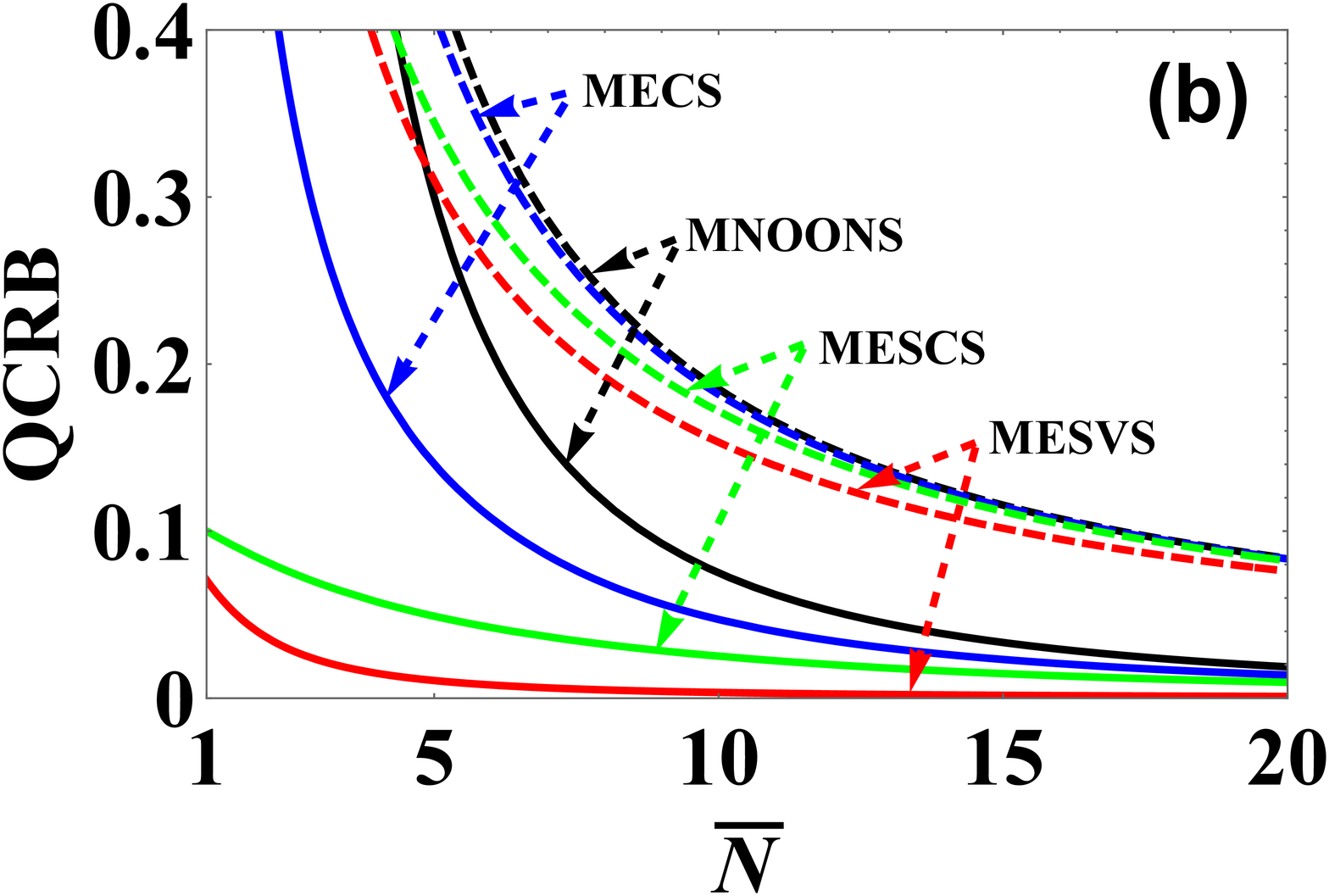} \newline
\caption{{}(Color online) The QCRB for the multiple angular displacement
estimation as a function of (a) the photon-loss strength $\protect \eta $
with $l=2,$ $d=15,$ and $\bar{N}=5,$ and of (b) the mean photon number $\bar{%
N}$ with $l=2,$ $d=15$ and $\protect \eta =0.7,$ when inputting the MNOONS
(black lines), the MECS (blue lines), the MESVS (red lines), and the MESCS
(green lines). The dashed and solid lines correspond to the photon-loss and
ideal cases, respectively.}
\end{figure}
\begin{figure}[tbp]
\label{Fig6} \centering \includegraphics[width=0.72\columnwidth]{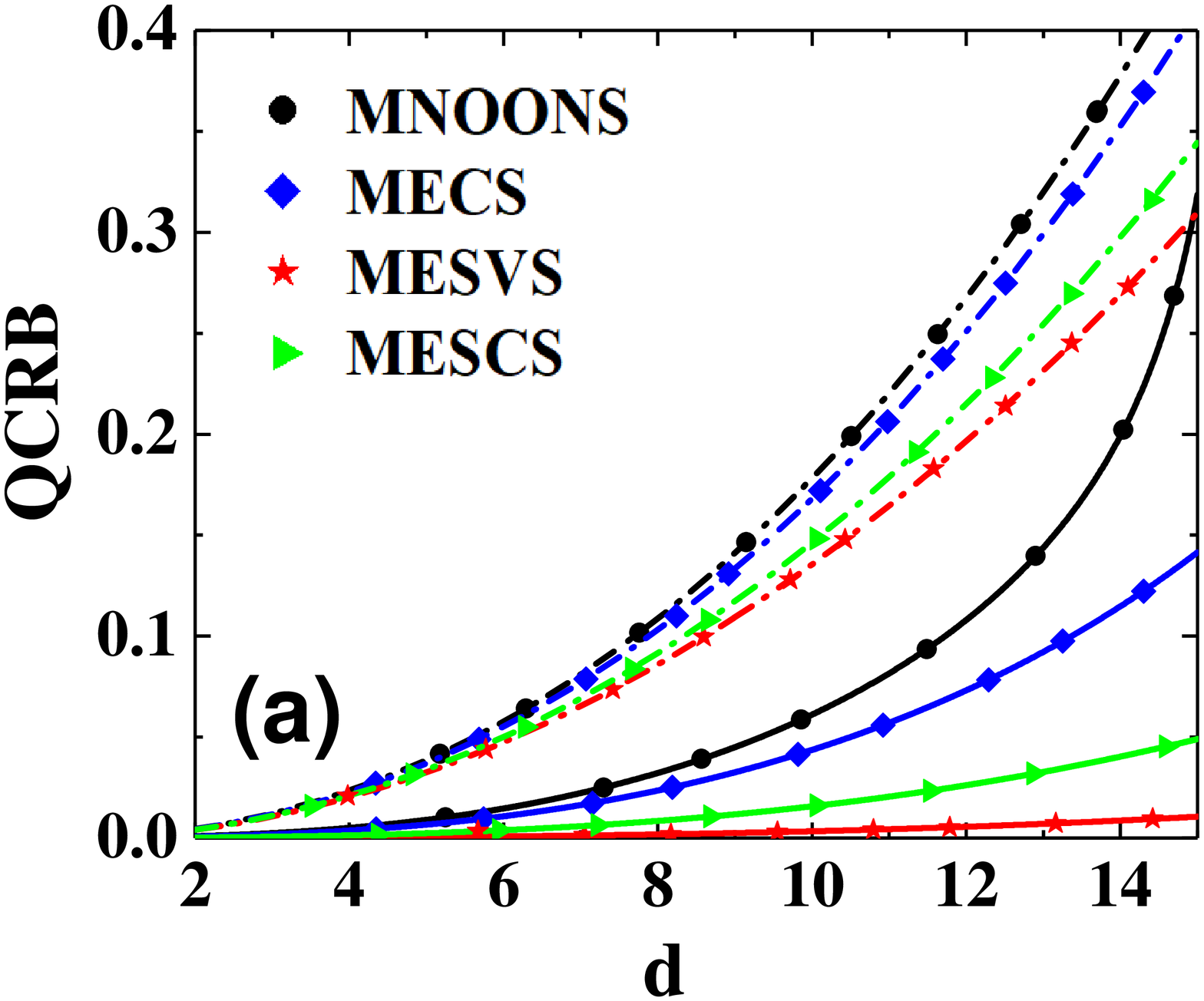}%
\newline
\includegraphics[width=0.72\columnwidth]{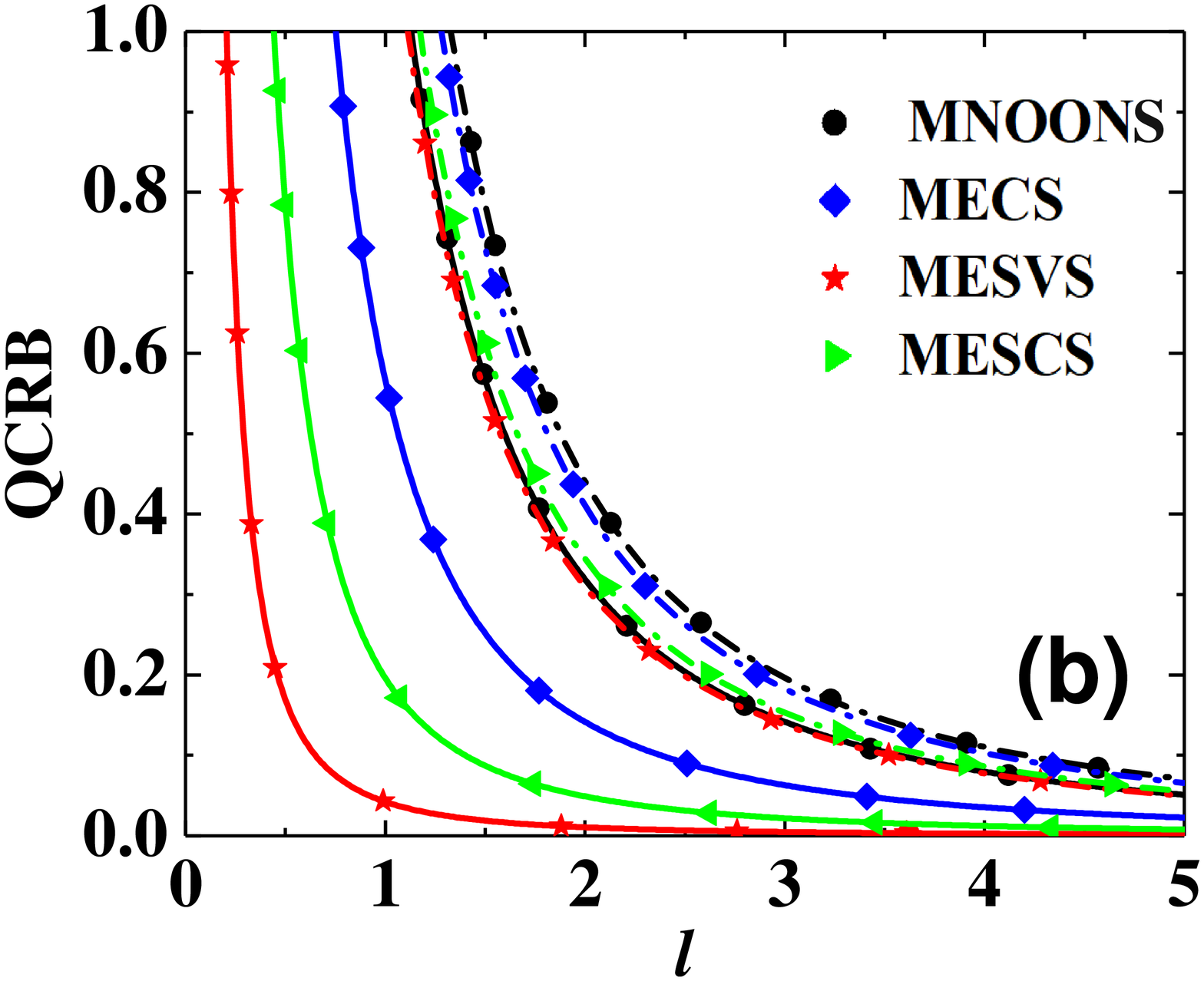} \newline
\caption{{}(Color online) The QCRB for the multiple angular displacement
estimation as a function of (a) the number of independent angular
displacements $d$ with $\protect \eta =0.7,$ $l=2$ and $\bar{N}=5$, and of
(b) the quanta number of the OAM $l$ with $\protect \eta =0.7,$ $d=15$ and $%
\bar{N}=5,$ when inputting the MNOONS (black lines), the MECS (blue lines),
the MESVS (red lines) and the MESCS (green lines). The dashed and solid
lines correspond to the photon-loss and ideal cases, respectively.}
\end{figure}
\begin{figure}[tbp]
\label{Fig7} \centering \includegraphics[width=0.72\columnwidth]{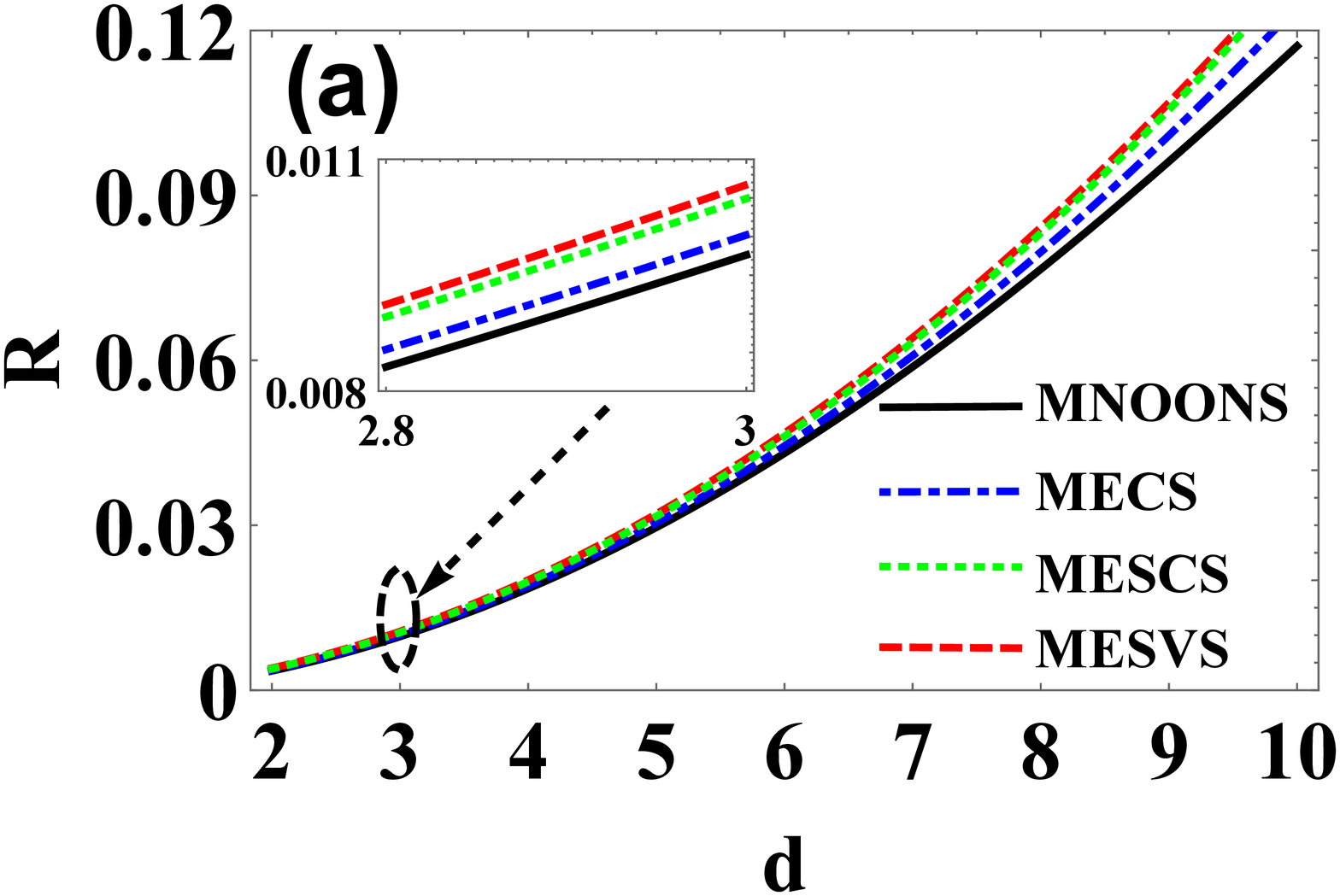}%
\newline
\includegraphics[width=0.72\columnwidth]{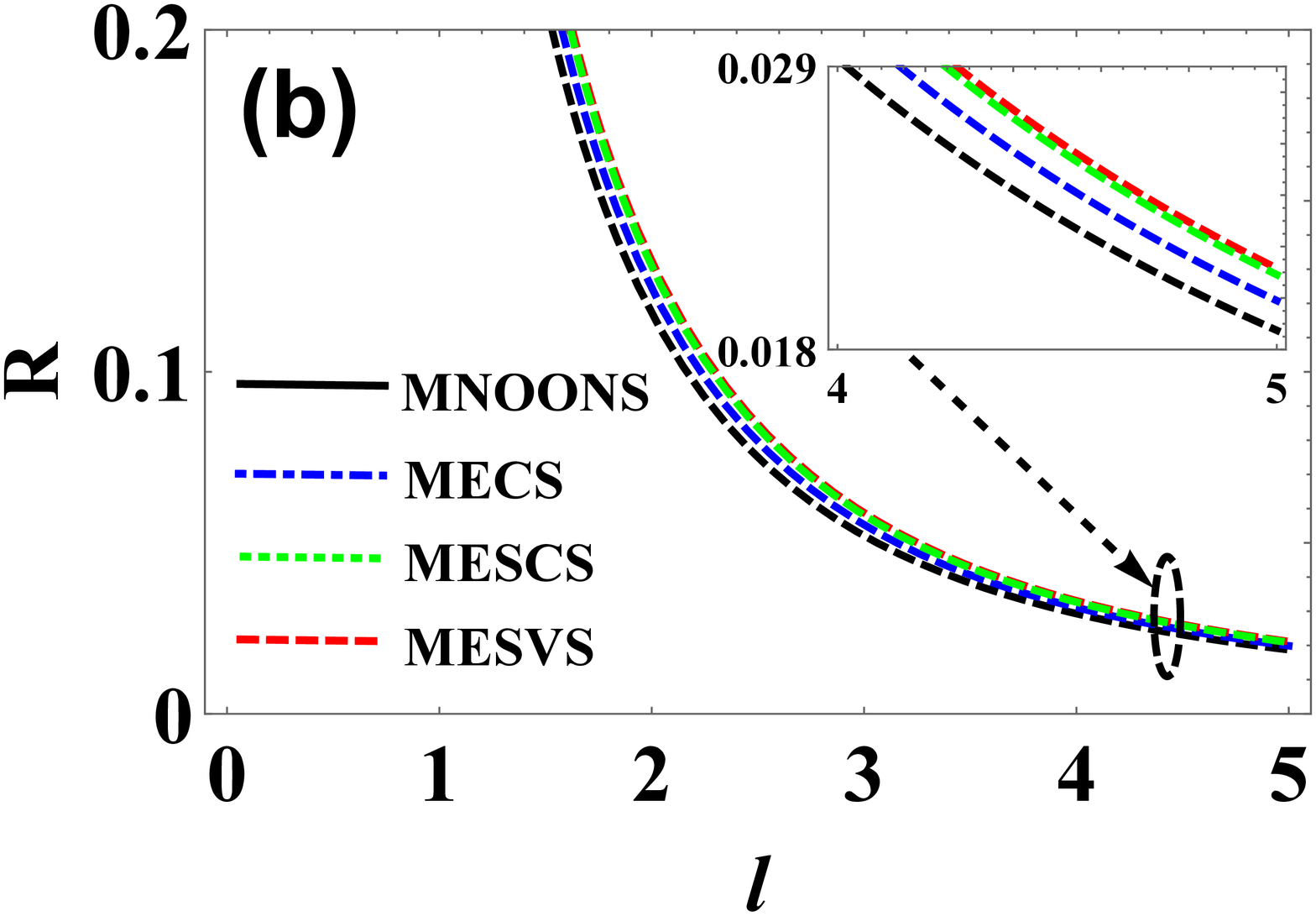} \newline
\caption{{}(Color online) The $R$ for the multiple angular displacement
estimation as a function of (a) the number of independent angular
displacements $d$ with $\protect \eta =0.7,$ $l=2$ and $\bar{N}=5,$ and of
(b) the quanta number of the OAM $l$ with $\protect \eta =0.7,$ $d=10$ and $%
\bar{N}=5,$ when inputting the MNOONS (black solid line), the MECS (blue
dot-dashed line), the MESVS (red dashed line) and the MESCS (green dot
line). }
\end{figure}

Next, in order to analyze the effects of the photon losses on the QCRB, let
us consider the four probe resources, involving the MNOONS $\left \vert \Psi
_{N}\right \rangle $, the MECS $\left \vert \Psi _{\alpha }\right \rangle $,
the MESVS $\left \vert \Psi _{r_{1}}\right \rangle $, and the MESCS $\left
\vert \Psi _{\beta ,r_{2}}\right \rangle $ [One can refer to Appendix B
about the expressions of the QCRB for these probe states]. When given the
values of $\bar{N}=5,$ $d=15$ and $l=2$, we plot the QCRB as a function of
the photon-loss strength $\eta $ for the four probe resources, as depicted
in Fig. 5(a). As we can see, the value of the QCRB for these probe states
increases rapidly with the decrease of $\eta $, implying that the accuracy
of the multiple angular displacement estimation is greatly affected by the
photon losses. In spite of this, the QCRB for the MESVS (red dashed line)
still shows the best performance even in the presence of photon losses,
followed by the MESCS, the MECS and the MNOONS. Moreover, in order to
compare the gap between the ideal and photon-loss cases, at fixed parameters
of $\eta =0.7,$ $d=15$ and $l=2$, we also show the QCRB changing with the
mean photon number $\bar{N}$ for the given probe resources, i.e., the MNOONS
(black lines), the MECS (blue lines), the MESVS (red lines), and the MESCS
(green lines), as pictured in Fig. 5(b). It is clearly seen that, although
the QCRB for the MNOONS performs worse than that for other probe states with
and without the photon losses, its gap between the ideal and photon-loss
cases is the smallest. This means that applying the MNOONS into multiple
angular displacement estimation systems is more robust against photon losses
than other probe resources at the same conditions. More interestingly, for
these probe resources, both the QCRB and the gap with and without the photon
losses can be further reduced as the mean photon number $\bar{N}$ increases,
implying that the increase of the mean photon number $\bar{N}$ of probe
states is a highly effective way to enhance the multiple angular
displacement estimation performance.

On the other hand, we also examine the influences of both $d$ and $l$ on the
QCRB under the photon losses when given parameters of $\eta =0.7$ and $\bar{N%
}=5$, as pictured in Fig. 6. Analogous to the ideal cases, for the photon
losses as the $d$ ($l$) increases, the multiple angular displacement
estimation precision becomes worse (more precise), and the MESVS still
maintains the highest estimation precision even beyond the ideal case of the
MNOONS. In addition, we also notice that, when given the same probe state,
e.g., the MESVS, as the $d$ ($l$) increases, the corresponding gap between
the ideal and photon-loss scenarios increases (decreases), implying that the
decrease of $d$ (or the increase of $l$) can not only improve the multiple
angular displacement estimation precision, but also enhance the robustness
against the photon losses. However, for different probe states, such as the
MESVS and the MECS shown in Fig. 6(a), their gaps can not be directly
visualized and compared. For this reason, to intuitively quantify and
visualize the gaps for the four probe states, we give the definition of the
robustness against the photon losses, i.e.,%
\begin{equation}
R=\left \vert \delta \theta \right \vert _{QCRB_{L}}^{2}-\left \vert \delta
\theta \right \vert _{QCRB}^{2}.  \label{18}
\end{equation}%
From Eq. (\ref{18}), the smaller the value of $R$, the smaller the gap
between $\left \vert \delta \theta \right \vert _{QCRB_{L}}^{2}$ and $%
\left
\vert \delta \theta \right \vert _{QCRB}^{2}$, meaning that the
robustness against the photon losses is stronger. To see this point, Fig. 7
shows the $R $ changing with $d$ and $l$ for the four probe resources when
given parameters of $\eta =0.7$ and $\bar{N}=5$. Visually, for the given
probe resources, the corresponding robustness $R$ is positively correlated
with $l$ and negatively correlated with $d.$ In particular, it is more
interesting that the MNOONS presents the best robustness, followed by the
MECS, the MESCS, and the MESVS, which is completely opposite to the
presentation accuracy of their multiple angular displacement estimations.
That is to say, the QCRB for\ the MNOONS shows the worst performance with
and without the photon losses compared to other probe resources, but the
usage of the MNOONS in the multiple angular displacement estimation systems
has the best robustness. Furthermore, it is worth mentioning that, when
comparing to other probe resources, the robustness performance for the MESVS
against the photon losses is relatively poor, but can gradually approach the
robustness of the MNOONS with the increase of $l$, which also means that the
OAM quantum number $l$ is profitably used for enhancing the robustness of
multiple angular displacement estimation systems.

\section{Conclusions}

In summary, we have revealed an important factor, i.e., the intramode
correlation of the probe state, which affects the multiple angular
displacement estimation precision with and without the photon losses. This
finding offers a reasonable explanation for the multiple angular
displacement estimation performance with $(d+1)$-mode NOON-like probe
states. The results show that the usage of the MESVS as the probe state is
more beneficial for obtaining the highest estimation precision than another
multimode probe state, which results from the intramode correlation of the
MESVS is the strongest. We have also considered the effects of the photon
losses on the multiple angular displacement estimation precision by the
means of the variational method.\textit{\ }The results suggest that the
accuracy of the multiple angular displacement estimation is greatly affected
by the photon losses, but the QCRB for the MESVS still shows the best
estimation performance when comparing to the one for another probe state.
More interestingly, different from the multiphase estimated systems, we can
also regulate and control the quanta number of the OAM $l$ to effectively
improve the robustness and precision of multiple angular displacement
estimation.\newline

\begin{acknowledgments}
We sincerely thank Prof. Yu-Ran Zhang and Heng Fan for helpful discussions.
This work was supported by the National Nature Science Foundation of China
(Grant Nos. 91536115, 11534008, 62161029); Natural Science Foundation of
Shaanxi Province (Grant No. 2016JM1005); Natural Science Foundation of
Jiangxi Provincial (Grant No. 20202BABL202002). Wei Ye is supported by both
the Natural Science Foundation of Jiangxi Province Youth Fund Project and
the Scientific Research Startup Foundation (Grant No. EA202204230) at
Nanchang Hangkong University.
\end{acknowledgments}

\noindent \textbf{Appendix\ A: The QCRB of four specific multimode entangled
state in the deal case}

Based on Eq. (\ref{7}), one can get the
\begin{align}
& \left \vert \delta \theta \right \vert _{QCRB(MNOONS)}^{2}  \notag \\
& =\frac{d}{16l^{2}(\bar{n}_{m\left( N\right) }^{2}g_{m\left( N\right)
}^{\left( 2\right) }+\bar{n}_{m\left( N\right) })}\left( 1+\frac{1}{%
g_{m\left( N\right) }^{\left( 2\right) }+\bar{n}_{m\left( N\right) }^{-1}-d}%
\right) ,  \notag \\
& \left \vert \delta \theta \right \vert _{QCRB(MECS)}^{2}  \notag \\
& =\frac{d}{16l^{2}(\bar{n}_{m\left( \alpha \right) }^{2}g_{m\left( \alpha
\right) }^{\left( 2\right) }+\bar{n}_{m\left( \alpha \right) })}\left( 1+%
\frac{1}{g_{m\left( \alpha \right) }^{\left( 2\right) }+\bar{n}_{m\left(
\alpha \right) }^{-1}-d}\right) ,  \notag \\
& \left \vert \delta \theta \right \vert _{QCRB(MESVS)}^{2}  \notag \\
& =\frac{d}{16l^{2}(\bar{n}_{m\left( r_{1}\right) }^{2}g_{m\left(
r_{1}\right) }^{\left( 2\right) }+\bar{n}_{m\left( r_{1}\right) })}\left( 1+%
\frac{1}{g_{m\left( r_{1}\right) }^{\left( 2\right) }+\bar{n}_{m\left(
r_{1}\right) }^{-1}-d}\right) ,  \notag \\
& \left \vert \delta \theta \right \vert _{QCRB(MESCS)}^{2}=\frac{d}{16l^{2}(%
\bar{n}_{m\left( \beta ,r_{2}\right) }^{2}g_{m\left( \beta ,r_{2}\right)
}^{\left( 2\right) }+\bar{n}_{m\left( \beta ,r_{2}\right) })}  \notag \\
& \times \left( 1+\frac{1}{g_{m\left( \beta ,r_{2}\right) }^{\left( 2\right)
}+\bar{n}_{m\left( \beta ,r_{2}\right) }^{-1}-d}\right) ,  \tag{A1}
\end{align}%
where we have set%
\begin{align}
\bar{n}_{m\left( N\right) }& =\breve{N}_{N}^{2}N,  \notag \\
g_{m\left( N\right) }^{\left( 2\right) }& =\frac{N-1}{\bar{n}_{m\left(
N\right) }},  \notag \\
\bar{n}_{m\left( \alpha \right) }& =\breve{N}_{\alpha }^{2}\alpha ^{2},
\notag \\
g_{m\left( \alpha \right) }^{\left( 2\right) }& =\frac{1}{\breve{N}_{\alpha
}^{2}},  \notag \\
\bar{n}_{m\left( r_{1}\right) }& =\breve{N}_{r_{1}}^{2}\sinh ^{2}r_{1},
\notag \\
g_{m\left( r_{1}\right) }^{\left( 2\right) }& =\frac{\breve{N}%
_{r_{1}}^{2}\left( 3\cosh 2r_{1}-7\right) \cosh ^{2}r_{1}+4}{2\bar{n}%
_{m\left( r_{1}\right) }^{2}},  \notag \\
\bar{n}_{m\left( \beta ,r_{2}\right) }& =\breve{N}_{\beta ,r_{2}}^{2}(\beta
^{2}+\sinh ^{2}r_{2}),  \notag \\
g_{m\left( \beta ,r_{2}\right) }^{\left( 2\right) }& =\frac{\breve{N}_{\beta
,r_{2}}^{2}(Z_{1}+Z_{2})+2}{\bar{n}_{m\left( \beta ,r_{2}\right) }^{2}},
\tag{A2}
\end{align}%
with%
\begin{align}
\breve{N}_{N}& =\frac{1}{\sqrt{1+d}},  \notag \\
\breve{N}_{\alpha }& =\frac{1}{\sqrt{\left( 1+d\right) (1+de^{-\alpha ^{2}})}%
},  \notag \\
\breve{N}_{r_{1}}& =\frac{1}{\sqrt{\left( 1+d\right) (1+d\sec hr_{1})}},
\notag \\
\breve{N}_{\beta ,r_{2}}& =\frac{1}{\sqrt{\left( 1+d\right) (1+de^{-\beta
^{2}(1-\tanh r_{2})}\sec hr_{2})}},  \notag \\
Z_{1}& =\beta ^{2}\sinh (2r_{2})+\left( 2\beta ^{2}-1\right) \cosh (2r_{2}),
\notag \\
Z_{2}& =\frac{3}{8}\cosh (4r_{2})+\beta ^{4}-2\beta ^{2}-\frac{11}{8}.
\tag{A3}
\end{align}

\noindent \textbf{Appendix\ B: The QCRB of four specific multimode entangled
state under the photon losses}

According to the Eq. (\ref{17}), one can obtain%
\begin{align}
& \left \vert \delta \theta \right \vert _{QCRB_{L}(MNOONS)}^{2}  \notag \\
& =\frac{d-1}{16l^{2}\bar{n}_{m\left( N\right) }}\left( \frac{1-\eta }{\eta }%
+\frac{1}{1+\bar{n}_{m\left( N\right) }g_{m\left( N\right) }^{\left(
2\right) }}\right) ,  \notag \\
& \left \vert \delta \theta \right \vert _{QCRB_{L}(MECS)}^{2}  \notag \\
& =\frac{d-1}{16l^{2}\bar{n}_{m\left( \alpha \right) }}\left( \frac{1-\eta }{%
\eta }+\frac{1}{1+\bar{n}_{m\left( \alpha \right) }g_{m\left( \alpha \right)
}^{\left( 2\right) }}\right) ,  \notag \\
& \left \vert \delta \theta \right \vert _{QCRB_{L}(MESVS)}^{2}  \notag \\
& =\frac{d-1}{16l^{2}\bar{n}_{m\left( r_{1}\right) }}\left( \frac{1-\eta }{%
\eta }+\frac{1}{1+\bar{n}_{m\left( r_{1}\right) }g_{m\left( r_{1}\right)
}^{\left( 2\right) }}\right) ,  \notag \\
& \left \vert \delta \theta \right \vert _{QCRB_{L}(MESCS)}^{2}  \notag \\
& =\frac{d-1}{16l^{2}\bar{n}_{m\left( \beta ,r_{2}\right) }}\left( \frac{%
1-\eta }{\eta }+\frac{1}{1+\bar{n}_{m\left( \beta ,r_{2}\right) }g_{m\left(
\beta ,r_{2}\right) }^{\left( 2\right) }}\right) ,  \tag{A4}
\end{align}%
where $\bar{n}_{m\left( N\right) },$ $g_{m\left( N\right) }^{\left( 2\right)
},$ $\bar{n}_{m\left( \alpha \right) },$ $g_{m\left( \alpha \right)
}^{\left( 2\right) },$ $\bar{n}_{m\left( r_{1}\right) },$ $g_{m\left(
r_{1}\right) }^{\left( 2\right) },$ $\bar{n}_{m\left( \beta ,r_{2}\right) },$
and $g_{m\left( \beta ,r_{2}\right) }^{\left( 2\right) }$ can be given by
Eq. (A2).

\end{document}